# A semi empirical compact equation of state for hard sphere fluids at any density

by Richard BONNEVILLE

Centre National d'Etudes Spatiales (CNES), 2 place Maurice Quentin, 75001 Paris, France

phone: 33 1 44 76 76 38, fax: 33 1 44 76 78 59, mailto:richard.bonneville@cnes.fr

**Abstract**

We propose a new semi empirical expression of the equation of state for a hard sphere fluid which is valid in the disordered phases over the whole density range below and above the freezing point. Starting from the existing numerical results for the virial coefficients, we elaborate a compact expression for the equation of state which is compatible both with the low and medium density behaviour (disordered stable phase) and with the asymptotic high density behaviour (disordered metastable phase). The resulting equation of state has a compact form and exhibits a simple pole at the close random packing density and a double pole at a density equal to 1. That equation of state only depends on the following quantities: the virial coefficients $B_2, B_3, B_4$, which can be exactly computed, the random packing density $\xi_0$, which is imposed by statistical geometry, and the residue of the pole in $\xi_0$. The results are in a fairly good agreement with the numerical data.

# 1. Introduction

The random packing of a very high number of identical hard spheres in a large container is a useful model for simple fluids: rare gases at high density or liquid metals are examples of such fluids; it is also a model for colloids, granular and amorphous materials. Consequently, an enormous quantity of theoretical [1, 2] as well as numerical [3, 4, 5, 6] and experimental [7] works have been dedicated to the equation of state of hard sphere fluids.

An essential parameter is the packing factor $\xi$, which is the ratio of the actual volume of a system of spheres to the total volume they occupy. The pioneering experiments on model systems performed by Bernal *et al.* [8] and Scott *et al.* [9] have shown that the random packing density for hard spheres is very sensitive to experimental conditions. $\xi$ can be increased up to a self-blocking situation ("loose random packing") whose precise value critically depends on the way the "fluid" has been poured into the container; by gently shaking or vibrating the container so as to simulate thermal agitation and to allow for local re-arrangements, the packing density comes closer to an upper limit $\xi_0$ ("close random packing") which is found to be about 0.637 [10, 11] when one extrapolates the experimental points to an infinite sample i.e. when the boundary effects become negligible.

Molecular dynamics and Monte Carlo-type calculations have confirmed those results but actually extensive numerical simulations have evidenced a more complex behaviour [12]: in the absence of any attractive term in the molecular interaction, there is no thermally induced liquid - gas phase transition but freezing of the fluid phase can occur when the packing density exceeds a value $\xi_f \cong 0.495$. Above $\xi_0$ only the ordered phase exists, its density between ultimately capped by the fcc packing density $\xi_{fcc} = \pi\sqrt{2}/6 \cong 0.7405$. Between $\xi_f$ and $\xi_0$ the fluid phase is metastable [13, 14]: a randomly packed fluid is very sensitive to shearing, in contrast with a regularly packed medium, and the close random packing limit $\xi_0$ can be exceeded if some shearing, locally creating ordered domains, is applied; the actual system then presents a mixture of ordered and disordered areas, the so-called "jammed states" [15, 16, 17]. Melting of the ordered phase occurs when the density goes down backwards below $\xi_m \cong 0.545$.

Quite generally the fluid pressure can be expressed as an expansion in powers of the density, the "virial expansion". For a system of $N$ identical hard spheres of diameter $D$ in a container of volume $V$ it can be written as

$$\frac{Pv}{k_B T} = \xi + 4\xi^2 + B_3 \xi^3 + B_4 \xi^4 + B_5 \xi^5 + B_6 \xi^6 + ... \tag{1}$$

P is the pressure, $v = \pi D^3 / 6$ is the molecular volume, $k_B$ is the Boltzmann constant, T is the absolute temperature and $\xi = Nv/V$ is the packing factor or reduced density. The number of particles N and the volume V are arbitrary large but finite, whereas $\xi$ is a given finite parameter. In this expansion, the first term is the ideal gas term, the second one the mean-field term, the third term and the subsequent ones account for the molecular correlations; $B_3 = 10$, $B_4 = 18.364768...$, the following coefficients can be computed numerically, the task getting increasingly harder with the order:

$$\frac{Pv}{k_B T} \cong \xi + 4\xi^2 + 10\xi^3 + 18.3648\xi^4 + 28.2245\xi^5 + 39.8151\xi^6 + 53.3444\xi^7 + ... \tag{2}$$

So far a large number of theoretical or semi-empirical equations of state, usually expressed as

$$\frac{PV}{Nk_B T} = F(\xi), \tag{3}$$

have been proposed for hard sphere systems [2]. In particular the famous Carnahan & Starling equation of state [18] is obtained by rounding the known virial coefficients as

$$\frac{PV}{Nk_B T} = 1 + 4\xi + 10\xi^2 + 18\xi^3 + 28\xi^4 + 40\xi^5 + 54\xi^6 ... \ . \tag{4a}$$

and extrapolating so that those coefficients appear to derive from a simple recursion

$$B_p = (p-1)(p+2) \tag{4b}$$

In compact form, the Carnahan & Starling equation of state is written as

$$\frac{PV}{Nk_B T} = \frac{1 + \xi + \xi^2 - \xi^3}{(1-\xi)^3} \tag{4c}$$

That expression is known to work remarkably well in the stable phase, i.e. for $\xi \lesssim 0.5$, and even a little beyond in the metastable phase [19]. Now it is obvious that at high density it does not work any longer since it has a triple pole divergence for $\xi = 1$, whereas the equation of state should exhibit a simple pole for $\xi = \xi_0$; actually it is known from the asymptotic expression of the virial coefficients [20] that the equation of state at high density takes the following form [21, 22]:

$$\frac{PV}{Nk_B T} \cong \frac{\lambda}{1 - \xi/\xi_0} \tag{5}$$

where $\lambda$ is a numerical coefficient whose value is about 2.8; in the high density limit the virial coefficients are then given by

$$\lim_{p\to\infty} B_p = \frac{\lambda}{(\xi_0)^{p-1}} \tag{6}$$

In the present paper, starting from the existing numerical results for the virial coefficients, we intend to exhibit a compact expression for the equation of state which is simultaneously compatible both with the low and medium density behaviour (disordered stable phase) and with the asymptotic high density behaviour (disordered metastable phase). We will see that our resulting equation of state only depends on the following quantities: the virial coefficients $B_2, B_3, B_4$, which can be exactly computed, the random packing density $\xi_0$, which is imposed by statistical geometry, and the residue of the pole in $\xi_0$.

## 2. Derivation of the virial coefficients

In order to account for the pole in $\xi = \xi_0$ it is necessary for the equation of state to have the following form:

$$\frac{PV}{Nk_BT} = 1 + 4\xi + 10\xi^2 \frac{f_1(\xi)}{1-\xi/\xi_0} \tag{7}$$

We develop $f_1(\xi)$ as a series of the density

$$\frac{PV}{Nk_BT} = 1 + 4\xi + 10\xi^2 \left( \frac{1 + a_4\xi + a_5\xi^2 + a_6\xi^3 + a_7\xi^4 + ...}{1-\xi/\xi_0} \right). \tag{8}$$

By construction, that equation of state is consistent with the low density limit $(\xi \to 0)$ and the high density limit $(\xi \to \xi_0)$. In particular it is known that for hard spheres [1,2] the compressibility factor $\frac{PV}{Nk_BT}$ and the contact value of the pair correlation function $g^{HS}(R=D_+)$ are connected through

$$\frac{PV}{Nk_BT} = 1 + 4\xi \, g^{HS}(R=D_+) \tag{9}$$

We see in the expression equ.(8) that we recover the expected low density behaviour

$$\lim_{\xi \to 0} \frac{PV}{Nk_BT} = 1 + o(\xi) \text{ and } \lim_{\xi \to 0} g(R=D_+) = 1 + o(\xi).$$

By identification of equ.(8) with the virial expansion we get

$$B_4/10 = (\xi_0)^{-1} + a_4$$
$$B_5/10 = (\xi_0)^{-2} + (\xi_0)^{-1} a_4 + a_5 \quad (10a)$$
$$B_6/10 = (\xi_0)^{-3} + (\xi_0)^{-2} a_4 + (\xi_0)^{-1} a_5 + a_6$$
...

or equivalently

$$B_5 = (\xi_0)^{-1} B_4 + 10 a_5$$
$$B_6 = (\xi_0)^{-1} B_5 + 10 a_6 \quad (10b)$$
$$B_7 = (\xi_0)^{-1} B_6 + 10 a_7$$
...

i.e. for $p \geq 4$

$$a_p = \left(B_p - (\xi_0)^{-1} B_{p-1}\right)/10 \quad (11)$$

Fig. 1 shows the virial coefficients $B_p$ derived from the numerical results by Clisby & Mc Coy for $p=1$ to $p=10$ [23] and from the results recently published by Wheatley for $p=11$ and $p=12$ [24]. If we consider those latter data, the curve $B_p(p)$ seems to exhibit an angular point for $p=11$. Now it is known that both the pressure $P(\xi)$ and its derivative $\frac{\partial P(\xi)}{\partial \xi}$ which is proportional to the isothermal compressibility are continuous in the transition from the low and medium density regime to the asymptotic regime; we thus believe that the numerical value for $B_{12}$ is doubtful. Nevertheless, the uncertainty attached to it is large ($\sigma \approx 26\%$); we have thus chosen to retain a corrected value at 1.3 $\sigma$ so as to suppress the angular point. Table 1 shows the $a_p$ coefficients, the inverse values $a_p^{-1}$ and the ratios $a_{p+1}/a_p$, for $p=4$ to $p=12$. The $a_p^{-1}$ decrease regularly in absolute value and seem to go to 0 whereas the ratio $a_{p+1}/a_p$ decreases to a constant value $\alpha \approx 1.30$. Consequently for p high enough the $a_p$ behave like $\alpha^p$, which suggests that $f_1(\xi)$ has a simple pole for $\xi = \alpha^{-1} \approx 0.769$, a value higher than the maximum possible density $\xi_{fcc} \cong 0.7405$; the gap is about 4%, one order of magnitude higher than the uncertainty given by Clisby & Mc Coy, thus too large enough not to be significant.

We henceforth will put $\alpha^{-1} \equiv \xi_1 \approx 0.769$ and then write:

$$f_1(\xi) = \frac{\beta}{1-\xi/\xi_1} + (1-\beta) + b_4 \xi + b_5 \xi^2 + b_6 \xi^3 + b_7 \xi^4 + ... \quad (12)$$

so that

$$a_4 = \beta(\xi_1)^{-1} + b_4$$
$$a_5 = \beta(\xi_1)^{-2} + b_5 \quad (13)$$
$$a_6 = \beta(\xi_1)^{-3} + b_6$$
...

i.e. for $p \geq 4$

$$b_p = a_p - \beta/(\xi_1)^{p-3} \quad (14)$$

Obviously if $p \to \infty$ then $b_p \to 0$ so that

$$\beta = \lim_{p \to \infty} a_p (\xi_1)^{p-3} \quad (15)$$

The $a_p$ coefficients and the derived $a_p(\xi_1)^{p-3}$ coefficients are displayed in Table 2a for $p = 4$ to $p = 12$ with $\xi_1 \cong 0.769$. The coefficients decrease in absolute value and seem to tend to a constant value $\approx -0.47$ so that we can write $\beta \approx -0.47$ if $\xi_1 \cong 0.769$. Table 2b shows the resulting values for the $b_p$ coefficients for $\beta \cong -0.47$ and $\xi_1 \cong 0.769$; it appears that for $p = 4$ to $p = 9$ the $b_p$ decrease with p in a nearly perfect linear fashion, i.e.

$$b_p \cong \mu p + \nu \quad (16)$$

and then smoothly go down to 0 for $p > 9$.

Given $\xi_1$ and $\beta$, and $a_4$ and $a_5$ coming from equ.(9b), $b_4$ and $b_5$ can be determined via equ.(12), and hence we get $\mu$ and $\nu$ as

$$\mu = b_5 - b_4$$
$$\nu = 5b_4 - 4b_5 \quad (17)$$

(the low order virial coefficients are a priori more accurate than the high order ones).

Henceforth we will treat $\xi_1$ and $\beta$ as adjustable independent parameters. For $\beta \cong -0.47$ and $\xi_1 \cong 0.769$ we get $\mu \cong -0.1444$ and $\nu \cong 1.4545$.

Let us truncate the $b_p$ sequence beyond $p = 9$, i.e. we put $b_p \cong 0$ for $p > 9$. Equ.(12) is then changed into

$$f_1(\xi) \cong \frac{\beta}{1 - \xi/\xi_1} + (1-\beta) + \nu\xi\left(1 + \xi + \xi^2 + \xi^3 + \xi^4 + \xi^5\right) + \mu\xi\left(4 + 5\xi + 6\xi^2 + 7\xi^3 + 8\xi^4 + 9\xi^5\right)$$

(18)

which after some manipulations becomes

$$f_1(\xi) \cong \frac{\beta}{1-\xi/\xi_1} + (1-\beta) + (\nu+3\mu)\xi\left(\frac{1-\xi^6}{1-\xi}\right) + \mu\xi\left(\frac{1-7\xi^6+6\xi^7}{(1-\xi)^2}\right) \tag{19}$$

We then obtain the following equation of state:

$$\frac{PV}{Nk_BT} = 1 + 4\xi + 10\xi^2\left(\frac{1}{1-\xi/\xi_0}\right)\left(\frac{\beta}{1-\xi/\xi_1} + (1-\beta) + (\nu+3\mu)\xi\left(\frac{1-\xi^6}{1-\xi}\right) + \mu\xi\left(\frac{1-7\xi^6+6\xi^7}{(1-\xi)^2}\right)\right) \tag{20}$$

## 3. Discussion

That equation of state describes the fluid behaviour in both the stable and the metastable disordered phases. It cannot account for the stable ordered phase since the equation of state of that phase must evidence a single pole at $\xi = \xi_{fcc}$; actually, the equation of state of the ordered phase is well represented by [25]

$$\frac{PV}{Nk_BT} \cong \frac{3}{1-\xi/\xi_{fcc}} \tag{21}$$

The following discussion will thus be restricted to the fluid phases.

The equation of state equ.(20) depends on the 2 parameters $\xi_1$ and $\beta$. The available computed values of the virial coefficients allow estimating the value of those 2 parameters. In order to refine that estimation, we remember that, from equ.(5) and equ.(20), we get for the residue $\lambda$ of the pole at $\xi_0$ the following expression

$$\lambda = 10\xi_0^2\left(\frac{\beta}{1-\xi_0/\xi_1} + (1-\beta) + (\nu+3\mu)\xi_0\left(\frac{1-\xi_0^6}{1-\xi_0}\right) + \mu\xi_0\left(\frac{1-7\xi_0^6+6\xi_0^7}{(1-\xi_0)^2}\right)\right) \tag{22}$$

With $\xi_1 \cong 0.769$ and $\beta \cong -0.47$, we would obtain $\lambda \cong -0.563$ which is absurd as we know that $\lambda$ is positive and $\lambda \approx 2.8$. Actually it appears that the numerical implementation of the formula equ.(22) shows a very high sensitivity of $\lambda$ to the value taken for $\xi_1$ and $\beta$. Treating $\xi_1$ and $\beta$ as adjustable independent parameters and exploring the parameter domain, we find that there is not a unique solution consistent with $\lambda \approx 2.8$ but a domain of possible solutions with $\lambda \approx 1$, as it is shown below for $\lambda \cong 2.765$, the value proposed by Speedy [21] and Robles et al. [22]:

| $\xi_1$ | $\beta$ |
|---------|---------|
| 0.97    | -0.272  |
| 1       | -0.361  |
| 1.03    | -0.458  |

Since it is not surprising to have a pole for $\xi = 1$ we retain the combination

$$\xi_1 \equiv 1, \beta \cong -0.361 \qquad (23)$$

It leads to $\mu \cong -0.3280$ and $\nu \cong 1.9386$ and to the equation of state below:

$$\frac{PV}{Nk_BT} = 1 + 4\xi + 10\xi^2 \left(\frac{1}{1-\xi/\xi_0}\right)\left(\frac{-0.361}{1-\xi} + 1.361 + 0.9547\xi\left(\frac{1-\xi^6}{1-\xi}\right) - 0.3280\xi\left(\frac{1-7\xi^6+6\xi^7}{(1-\xi)^2}\right)\right) \qquad (24)$$

(N.B.: A different choice for $\lambda$ while keeping $\xi_1 \equiv 1$ would have given a different value for $\beta$ and thus for the coefficients $\mu$ and $\nu$ although in all cases $\nu + 3\mu \approx 1$ and $\nu \approx 2$.)

Equ.(24) can be re-written as:

$$\frac{PV}{Nk_BT} = 1 + 4\xi + 10\xi^2 \left(\frac{1}{1-\xi/\xi_0}\right)\left(\frac{(1-1.7343\xi + 0.4063\xi^2) + (1.3413\xi^7 - 1.0133\xi^8)}{(1-\xi)^2}\right) \qquad (25)$$

With $\xi < 0.7$ the second bracket in the numerator of equ.(25) is at least one of order of magnitude smaller than the first one; if we neglect it we finally get

$$\frac{PV}{Nk_BT} \cong 1 + 4\xi + 10\xi^2 \left(\frac{1}{1-\xi/\xi_0}\right)\left(\frac{1-1.7343\xi + 0.4063\xi^2}{(1-\xi)^2}\right) \qquad (26)$$

That simplified expression exhibits a simple pole for $\xi = \xi_0$ and a double pole for $\xi = 1$. Note that the 4th virial coefficient derived from equ.(26) is $10 \times (0.637^{-1} - 1.7343 + 2) = 18.365$, i.e. the exact value as expected.

Noting that $0.4063 = 0.6374^2 \cong \xi_0^2$ it is tantalizing to write

$$\frac{PV}{Nk_BT} \cong 1 + 4\xi + 10\xi^2 \left(\frac{1}{1-(\xi/\xi_0)}\right)\left(\frac{1-1.7343\xi + (\xi \times \xi_0)^2}{(1-\xi)^2}\right)$$

However it may well be a numerical coincidence.

Our semi empirical equation of state equ.(26) only depends on the following quantities: the virial coefficients $B_2, B_3, B_4$, which can be exactly computed, the random packing density $\xi_0$ and the parameter $\lambda$, i.e. the residue of the pole in $\xi_0$. The random close packing limit is imposed by statistics and geometry only, not by thermodynamics, even if a complete molecular theory should be compliant with it; the value 0.637 (Scott et al. and Finney) is quite close to $2/\pi$ [12,20] but this has not been demonstrated to date. $\lambda$ has been given the value 2.765 proposed by Speedy and Robles et al. as previously said. In summary, our equation of state connects with a minimum apparatus the two types of regimes where there is a sound theoretical basis: the low density regime and the asymptotic high density regime.

The fact that $\xi_1$ is significantly above 0.77 means that the behaviour of the high order virial coefficients (beyond $p=12$) changes with respect to the behaviour of the low order coefficients; actually we had shown in [20] that the transition between the low and medium density regime well described by the Carnahan & Starling equation of state and the high density regime described by the asymptotic equation of state would occur at $p=13$ or $p=14$. We had also shown in [20] that the transition between the low and medium density regime and the high density regime is smooth and occurs around the freezing density.

Fig. 2a shows a graphical comparison covering the whole concentration range of the fluid phase between the Carnahan & Starling equation of state equ.(4), Speedy's asymptotic equation of state equ.(5) and our new equation of state equ.(26). Fig.2b is a zooming onto the low and medium density phase $(0 \leq \xi \leq 0.50)$ and Fig.2c is a zooming onto the high density phase $(0.50 \leq \xi \leq 0.637)$, the transition between the stable phase and the metastable phase occurring at the freezing density $\xi_f \cong 0.495$. It appears that in the low and medium density region our curve and Carnahan & Starling's one are hardly distinguishable and that in the high density region our curve and the asymptotic regime one are hardly distinguishable. We have not included in the graphics a comparison with the results of the available numerical simulation works, e.g. from Wu and Sadus [6], so as not to overload the figures as the curves would not be distinguishable. In addition, Fig.3 shows a graphical comparison of the contact value of the pair correlation function $g^{HS}(R=D_+)$ as derived from the equation of state via equ.(9) between the Carnahan & Starling equation of state equ.(4), Speedy's asymptotic equation of state equ.(5) and our new equation of state equ.(26) over the whole concentration range of the fluid phase.

Besides the simple Carnahan & Starling formula, numerous equations of state have been proposed for hard sphere fluids, many of them involving a more or less large set of parameters fitted with respect to the molecular dynamics calculations and/or to the numerically computed virial coefficients, e.g. via a Padé approximant approach (see in ref.[2] chapter 3 by Mulero et al., and more recently [26, 27, 28, 29]). A wide fraction of those equations of state is focused on modelling the stable fluid phase below the freezing density and does not account for the close random packing singularity. Quite recently a rather simple and compact equation of state covering the whole density range up to $\xi_0$ has been proposed by a modification of the Carnahan & Starling equation of state [30]. The latter can be derived from the following free energy F:

$$\frac{F}{Nk_BT} = \frac{3\xi}{1-\xi} + \frac{\xi}{(1-\xi)^2} \qquad (27)$$

In ref.[30] the above expression is tentatively replaced by

$$\frac{F}{Nk_BT} = \frac{3\xi}{1-\xi} + \frac{\xi}{(1-\xi)^\alpha (1-\Phi)^{2-\alpha}} \qquad (28a)$$

together with

$$\Phi = \xi\left(1 + \left(\frac{1-\xi_0}{\xi_0}\right)\exp(\beta(\xi-\xi_0))\right) \qquad (28b)$$

$\alpha$, $\beta$ and $\xi_0$ are supposed to be universal parameters; fitting those 3 quantities with respect to the molecular dynamics results of [4] and [5] gives $\alpha \cong 1$, $\beta \cong 50$ and $\xi_0 \cong 0.655$. If $\xi \to 0$ then $\Phi \cong \xi$ and the above system of equations gives back the Carnahan & Starling equation of state, whereas if $\xi \to \xi_0$ then $\Phi \cong \xi/\xi_0$ and the equation of state has a singularity for $\xi = \xi_0$. Although that equation of state shows good numerical results it predicts a double pole for the singularity in $\xi_0$ whereas a simple pole is expected and the fitted value for the random close packing limit is too high.

## 4. Conclusion

We have worked out a new compact semi empirical expression of the equation of state for a hard sphere fluids which is valid in the disordered phases over the whole density range below (stable phase) and above (metastable phase) the freezing point. The results are in a fairly good agreement with the available numerical data.

# Table captions

**Table 1:**

The $a_p$ coefficients, their inverse values $a_p^{-1}$ and the ratios $a_{p+1}/a_p$ for $p=4$ to $p=12$; figures from Clisby & Mc Coy for $p=4$ to $p=10$ and from Wheatley for $p=11$ and $p=12$ (modified value for $p=12$).

**Table 2:**

**Table (2a):** The $a_p(\xi_1)^{p-3}$ coefficients for $p=4$ to $p=12$ with $\xi_1 \cong 0.769$.

**Table (2b):** The $b_p$ coefficients for $p=4$ to $p=12$ with $\xi_1 \cong 0.769$ and $\beta \cong -0.47$.

**Table 1**

| p | $B_p$ | $a_p$ | $1/a_p$ | $a_{p+1}/a_p$ |
|---|---|---|---|---|
| 4 | 18.364768 | 0.2657 | 3.7639 | -0.2344 |
| 5 | 28.2245 | -0.0623 | -16.0566 | 7.2573 |
| 6 | 39.8151 | -0.4520 | -2.2125 | 2.0348 |
| 7 | 53.3444 | -0.9197 | -1.0873 | 1.6588 |
| 8 | 68.5375 | -1.5256 | -0.6555 | 1.4320 |
| 9 | 85.8128 | -2.1846 | -0.4578 | 1.3284 |
| 10 | 105.7751 | -2.9019 | -0.3446 | 1.3172 |
| 11 | 127.9263 | -3.8225 | -0.2616 | 1.3068 |
| 12 | 150.9949 | -4.9951 | -0.2002 | |

**Table 2a**

| p | $a_p (\xi_1)^{p-3}$ |
|---|---|
| 4 | 0,2043 |
| 5 | -0,0368 |
| 6 | -0,2055 |
| 7 | -0,3216 |
| 8 | -0,4103 |
| 9 | -0,4518 |
| 10 | -0,4615 |
| 11 | -0,4675 |
| 12 | -0,4698 |

**Table 2b**

| p | $b_p$ |
|---|---|
| 4 | 0,8769 |
| 5 | 0,7325 |
| 6 | 0,5815 |
| 7 | 0,4243 |
| 8 | 0,2221 |
| 9 | 0,0881 |
| 10 | 0,0534 |
| 11 | 0,0207 |
| 12 | 0,0025 |

**Figure captions**

**Figure 1:**

The virial coefficients $B_p(p)$, data from Clisby & Mc Coy for $p = 1$ to $p = 10$ and from Wheatley for $p = 11$ and $p = 12$.

**Figure 2:**

**2a:** Graphical comparison over the whole concentration range of the fluid phase between the Carnahan & Starling equation of state, Speedy's asymptotic equation of state and our new equation of state.

**2b:** Zooming onto the low and medium density phase $(0 \leq \xi \leq 0.50)$.

**2c:** Zooming onto the high density phase $(0.50 \leq \xi \leq 0.637)$.

**Figure 3:**

**2a:** Graphical comparison of the contact value of the pair correlation function $g^{HS}(R=D_+)$ over the whole concentration range of the fluid phase between the Carnahan & Starling equation of state, Speedy's asymptotic equation of state and our new equation of state.

**Fig.1&**

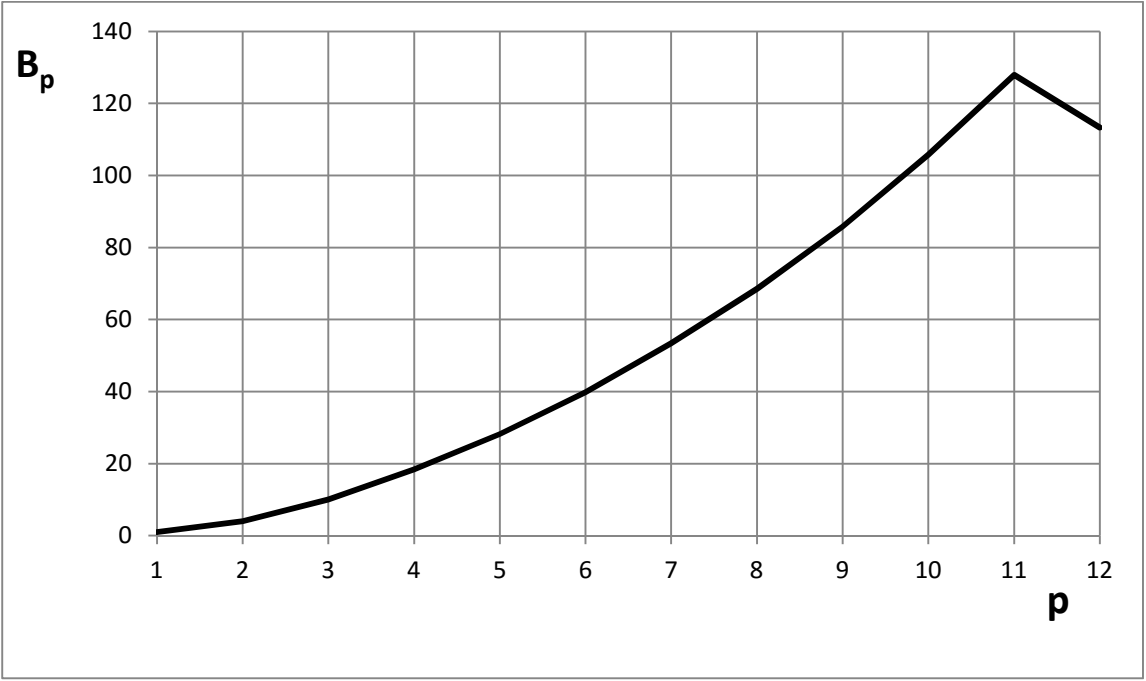

**Fig.2a**

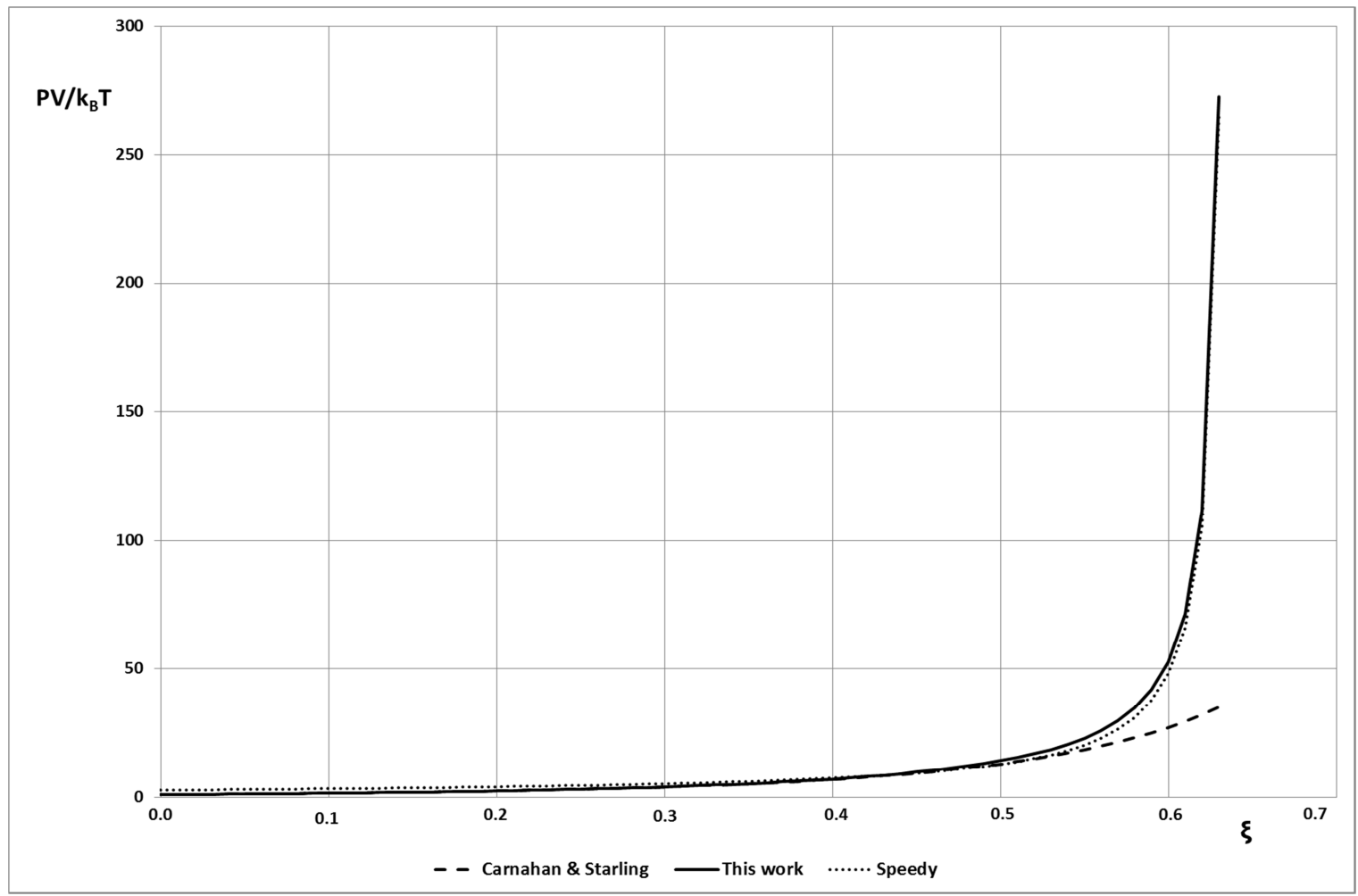

**Fig 2b**

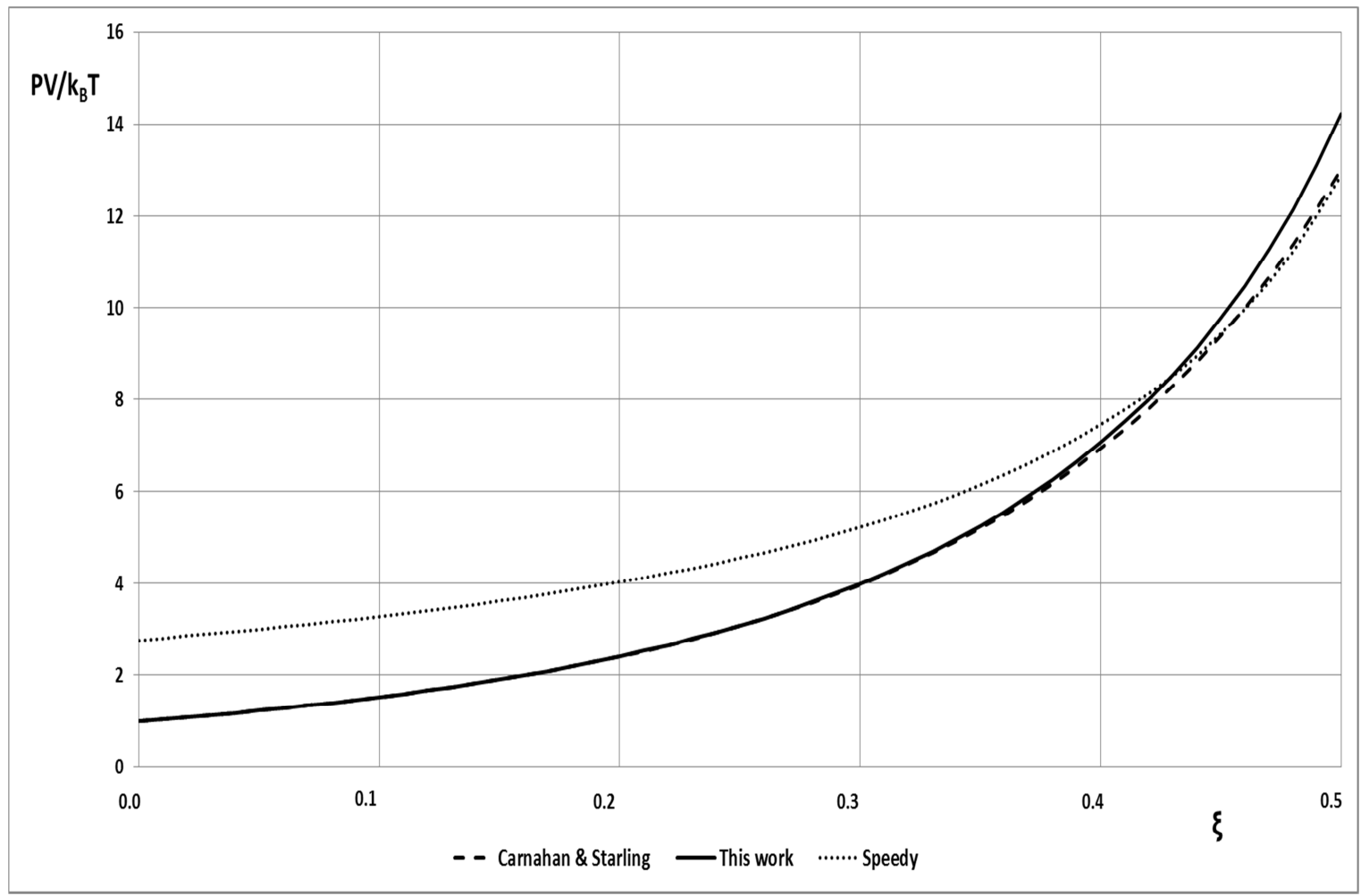

**Fig2c**

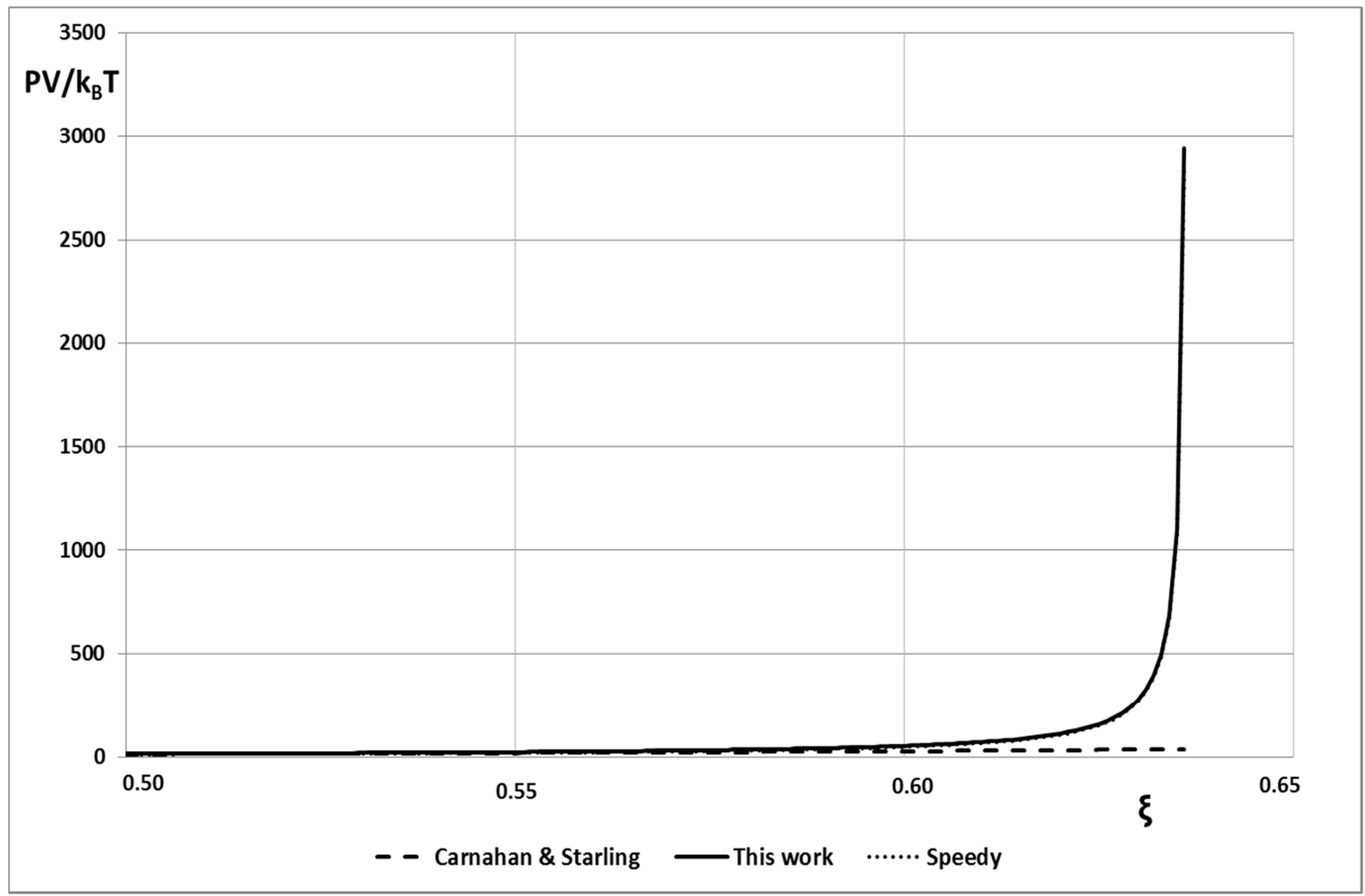

**Fig3**

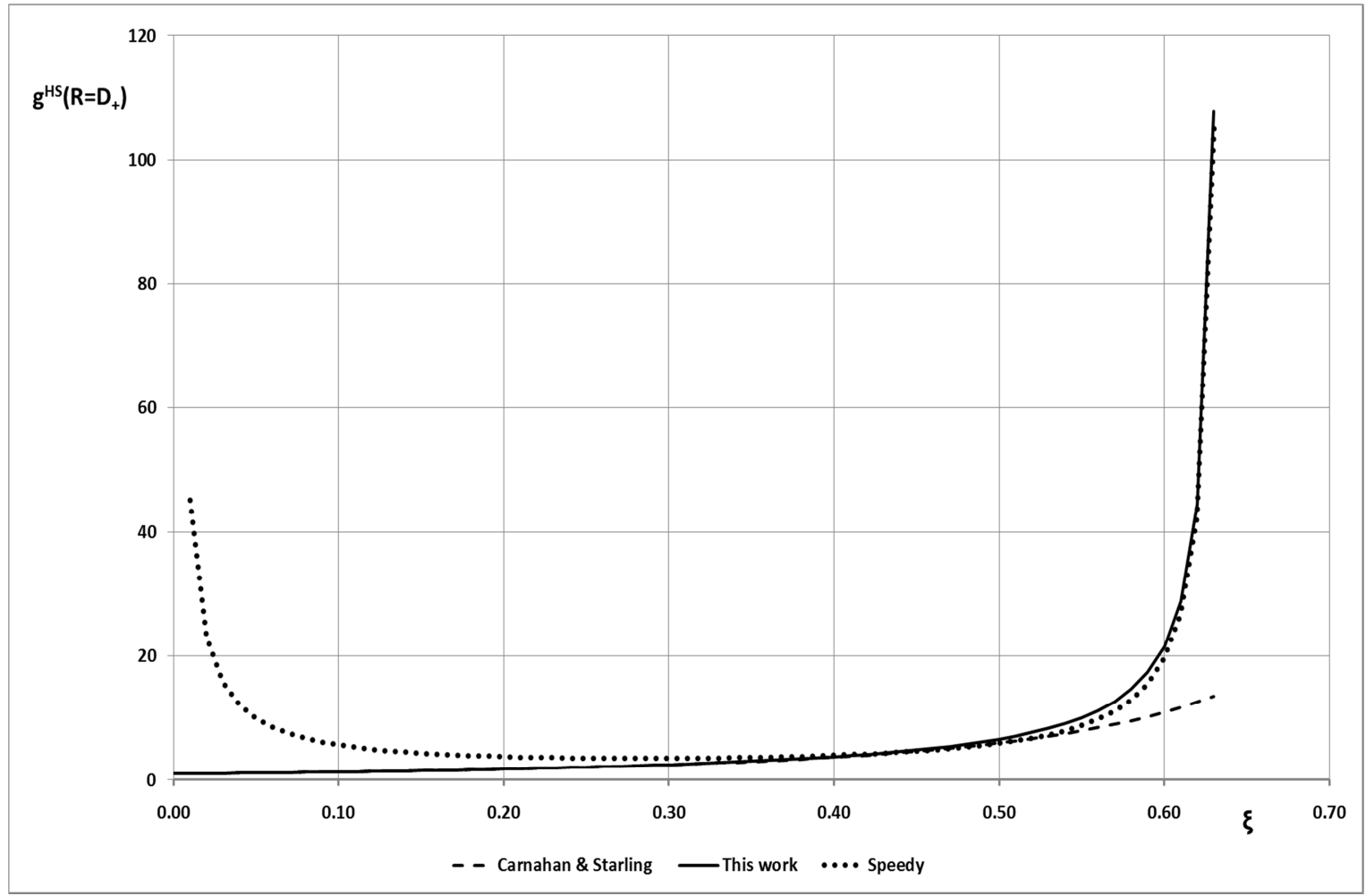